\documentclass[twoside]{article}
\usepackage[dvips]{graphicx}
\usepackage{fleqn,espcrc2}

\newcommand{\AmS}{{\protect\the\textfont2
  A\kern-.1667em\lower.5ex\hbox{M}\kern-.125emS}}
\providecommand{\openone}{\leavevmode\hbox{\small1\kern-3.5pt\normalsize1}}
\title{Order parameter of the chiral symmetry breaking Green Functions at one loop\thanks{Talk given at the High--Energy Physics International
Conference on Quantum Chromodynamics, 3-7 July (2006), Montpellier (France);
IFIC/06$-$32 FTUV/06$-$1019 report. To appear in the Proceedings.}
}

\author{V.~Mateu\address{Departament de F\'\i sica Te\`orica, IFIC, Universitat de Val\`encia - CSIC 
\\ Apt. Correus 22085, E-46071 Val\`encia, Spain }}
\begin{document}

\begin{abstract}\noindent
Since order parameter of the chiral symmetry breaking Green Functions are a useful link between the OPE expansion and $\chi$PT we perform a calculation to the NLO in $\alpha_s$, working in the chiral limit, for all the 2 and 3 point ones. These Green Functions have no perturbative term in the chiral limit and thence their main contribution comes from the $\left\langle \bar{q}q \right\rangle $ operator. 

\end{abstract}
\maketitle
\section{Introduction}\noindent
Although nowadays QCD is regarded as the fundamental theory responsible for all the strong interaction phenomenology, we are still far from a sheer understanding of many of its features.


Within the light quark sector and in the low energy regime (E $<< m_\rho$) QCD exhibits a spontaneous breakdown of the chiral symmetry. 
It is believed that the operator responsible for this phenomenon is the quark condensate $\left\langle \bar{q}q \right\rangle $, which acquires a non-vanishing value. 
Phenomenologically, this feature is reflected  through the appearance in the spectrum of the so called pseudo-Goldstone bosons. The effective theory governing their dynamics is Chiral Perturbation Theory ($\chi$PT) \cite{Chiral}. This theory consists in an expansion of the physical observables in terms of the momentum and quark masses.

In the deep Euclidean region ($q^2<<- m_\rho^2$) one can use the Operator Product Expansion (OPE) \cite{OPE}, which is a series in inverse powers of momentum, each term multiplied by the appropriate QCD operator.

For the intermediate region there is only one known expansion parameter, namely $1/N_C$ \cite{LargeN}. Large $N_C$ tells us that this region is dominated by the hadronic resonances, that dictate their dynamics and implies that there must be an infinite number of them. All these features can be put together in a Lagrangian formalism known as the Resonance Chiral Theory (R$\chi$T) \cite{Ecker:1988te}. Still we lack a method to handle an infinite tower of them, and then the so called minimal hadronic ansatz (MHA) is usually used, meaning that we only consider the lowest lying ones. One is then able to perform a matching of the three regimes and thence to estimate the values of the low energy constants (LEC's) of the $\chi$PT Lagrangian \cite{matching}. Some effort has been done to push this study towards $1/N_C$ corrections through resonance loops \cite{res1loop} and in the direction of including more resonances in the ansatz \cite{Sanz-Cillero:2005ef}.

Order parameter Green Functions are those which in the chiral limit have no 
purely perturbative term in the OPE 
and thus their leading contribution results from the  $\left\langle \bar{q}q \right\rangle $ operator. Since this operator is the responsible for the chiral symmetry breaking those Green Functions encode essential information for a better understanding of confinement.

So far only the $\mathcal{O}(\alpha_s^0)$ has been calculated. In this work we perform the first gluonic corrections \cite{mateu-jamin-pich} which translates into scale and logarithm dependences. This might be useful for a scale dependence determination of the LEC's and for including the full tower of resonances. 

\section{Definitions}
\noindent 
The QCD currents are colour singlets 
quarks bilinears. In general they can be written as:
\begin{equation}
J^{a}(x)  =  :\bar{q}(x)\Gamma_J\left( \frac{\lambda^{a}}{2}\right) q(x):\,,
\end{equation} 
where $\lambda^{a}$ are the Gell-Mann matrices of the SU(3)
group in flavour space and $\Gamma_J$ is a Dirac matrix.

The mathematical definition of the Green functions is as 
follows:
\begin{eqnarray}
\Pi_{123}& \!\!\!\!\!\!= &\!\!\!\!\! i^{2}\int\!\!\mathrm{d}^{4}x\,\mathrm{d}^{4}y\, e^{i(qx+py)}\!\!\left\langle T\{\! J_{1}(x)J_{2}(y)J_{3}(0)\}\right\rangle\!, \nonumber\\
\Pi_{12}& \!\!\!\!\!\!= &\!\!\!\!\!  i\int\!\!\mathrm{d}^{4}x\,e^{ipx}\left\langle T\{ J_{1}(x)J_{2}(0)\}\right\rangle\,,  \label{eq:def-3}\end{eqnarray} 
The expectation value in eq.~(\ref{eq:def-3}) is to
be understood with respect to the non-perturbative QCD vacuum
. This vacuum is responsible for the breakdown of the chiral
symmetry. 
%

The OPE master formula expands the Green Functions into an infinite
sum of 
local operators where the dependences in the space-time coordinates are enclosed in the
Wilson coefficients:
\begin{eqnarray}
\lim_{p\rightarrow\infty}\left\langle \mathcal{O}_{1}\mathcal{O}_{2}\right\rangle  & = & \sum_{i=1}^{\infty}C_{i}\left(p,q,\mu;\alpha_{s}\right)\left\langle \mathcal{O}_{i}(\mu,0)\right\rangle \,,\nonumber\\
C_{i}\left(p,\mu;\alpha_{s}\right) & = & \sum_{j=0}^{\infty}\left(\frac{\alpha_{s}}{\pi}\right)^{j}C_{ij}\left(p,\mu\right)\,,\\
\mathcal{O}_{1} & = & \openone\;\mathrm{(identity)}\,,\nonumber\\
\mathcal{O}_{2} & = & \left\langle \bar{q}q\right\rangle\,,\nonumber \\
 & \cdots\nonumber\end{eqnarray}
In this work we deal with functions with $C_{1}=0$ to all orders
in $\alpha_{s}$. These are called order parameteres of the chiral symmetry breaking.
We will concentrate on the $\alpha_{s}$ correction to the Wilson
coefficient $C_{2}$.

The Vector and Axial-Vector currents are conserved in the chiral limit. This fact is reflected in the Green functions through the Ward identities. The Ward identity which the $\left\langle AP\right\rangle $ correlator must follow fixes completely in a non-perturbative way its structure \cite{mateu-jamin-pich}:
\begin{eqnarray}
\left(\Pi_{AP}^{\mu}\right)^{ab} & = & 2i\delta^{ab}\frac{\left\langle \bar{q}q\right\rangle }{q^{2}}q^{\mu}\,,\end{eqnarray}
and so does not get any $\alpha_s$ correction. The only hadron interchanged by the two currents is a pseudo-Goldstone boson.

\section{Two point functions}
\begin{figure}[tbh!]
\begin{center}\includegraphics[%
  scale=0.5]{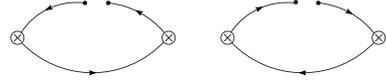}\end{center}\vspace*{-0.9cm}
\caption{$\mathcal{O}(\alpha_{s}^{0})$ contributions to the quark condensate
operator \label{cap:2-point-tree}}\end{figure}

\noindent
Two point functions are of special interest since at tree level they only
receive contributions from single Goldstone and resonance exchange. 
Let us concentrate in the $\left\langle VT\right\rangle $ correlator, which except for the $\left\langle AP\right\rangle $ is the only two point order parameter of the chiral symmetry breaking. Vector meson resonances are interchanged between the two currents. From general principles it can be inferred that:
\begin{equation}
\left(\Pi_{VT}\right)_{\mu\nu\alpha}^{ab}(p)= \Pi_{VT}\left(p^{2}\right)\delta^{ab}\left(p_{\alpha}g_{\mu\nu}-p_{\nu}g_{\mu\alpha}\right)\,,
\end{equation} 
Fig.~\ref{cap:2-point-tree} shows all the diagrams entering the $\mathcal{O}(\alpha_{s}^{0})$
for a two point correlator 
and Fig.~\ref{cap:2-point-tree} their first order gluonic corrections.\vspace*{-0.4cm}
\vspace*{-1.3cm}
\begin{figure}[tbh!]
\begin{center}\includegraphics[%
  scale=0.4]{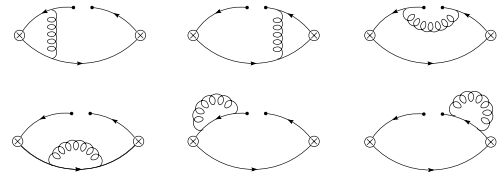}\end{center}\vspace*{-0.9cm}
\caption{Gluonic corrections to Fig.~\ref{cap:2-point-tree}\label{cap:2-point-Gluonic}}
\end{figure}\vspace{-0.5cm}

In Fig.~\ref{cap:2-point-Gluonic} one observes there are gluons attached to quark propagators
with zero incoming momenta and this may cause infrared divergences. In order to regularize
these divergences we use dimensional regularization. Doing so, one cannot distinguish between 
ultraviolet and infrared ones, all of them manifest as $1/\hat{\epsilon}$ (and $\log(\mu)$)
and hence the total divergence of a diagram is the sum of both.  In this scheme only the box
diagram has an infrared divergence. The divergence of this diagram is $\frac{1}{\hat{\epsilon}}\frac{(3+a)}{4}\frac{\alpha_{s}}{\pi}\frac{C_{F}}{2}$
where $a$ is the arbitrary gauge parameter of QCD\footnote{We perform all our calculations in an arbitrary gauge $a$.}. Together with the tree level amplitude we get the structure:
\begin{eqnarray}
\left\langle \bar{q}q\right\rangle \left(1-\frac{1}{\hat{\epsilon}}\frac{(3+a)}{4}\frac{\alpha_{s}}{\pi}C_{F}\right)=\left\langle \bar{q}_{R}q_{R}\right\rangle Z_{m}Z_{2F}\\=\left\langle \bar{q}_{B}q_{B}\right\rangle Z_{m}=\frac{m_{B}}{m_{R}(\mu)}\left\langle \bar{q}_{B}q_{B}\right\rangle =\left\langle \bar{q}q\right\rangle _{R}(\mu)\,,\nonumber
\end{eqnarray} 
where $q_{B}=Z_{2F}^{1/2}q_{R}$, $m_{B}=Z_{m}m$ and the subscript
$R$ stands for renormalized and $B$ for bare. We have used the fact
that the product $m\left\langle \bar{q}q\right\rangle $ is a renormalization
group invariant quantity and so the divergence is absorbed in the
renormalization of the condensate.

After renormalization the final result reads  \cite{mateu-jamin-pich}:

\begin{equation}
\Pi_{VT}=-i\frac{\left\langle \bar{q}q\right\rangle}{p^{2}}\!\left\{ 1\!-\!\frac{\alpha_{S}}{\pi}C_{F}\!\!\left[\log\left(-\frac{p^{2}}{\mu^{2}}\right)-1\right]\right\}\label{VTresult},
\end{equation} 
which is independent of the $a$ parameter, required by gauge invariance and it is also a good check for our calculation.

\section{Three point functions}
\begin{figure}[tbh!]
\begin{center}\includegraphics[%
  scale=0.5]{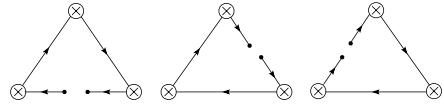}\end{center}\vspace*{-0.9cm}
\caption{$\mathcal{O}(\alpha_{s}^{0})$ contributions to the quark condensate
operator \label{cap:3-point-tree}}
\end{figure}

\begin{figure}[tbh!]\vspace*{-0.5cm}
\begin{center}\includegraphics[%
  scale=0.4]{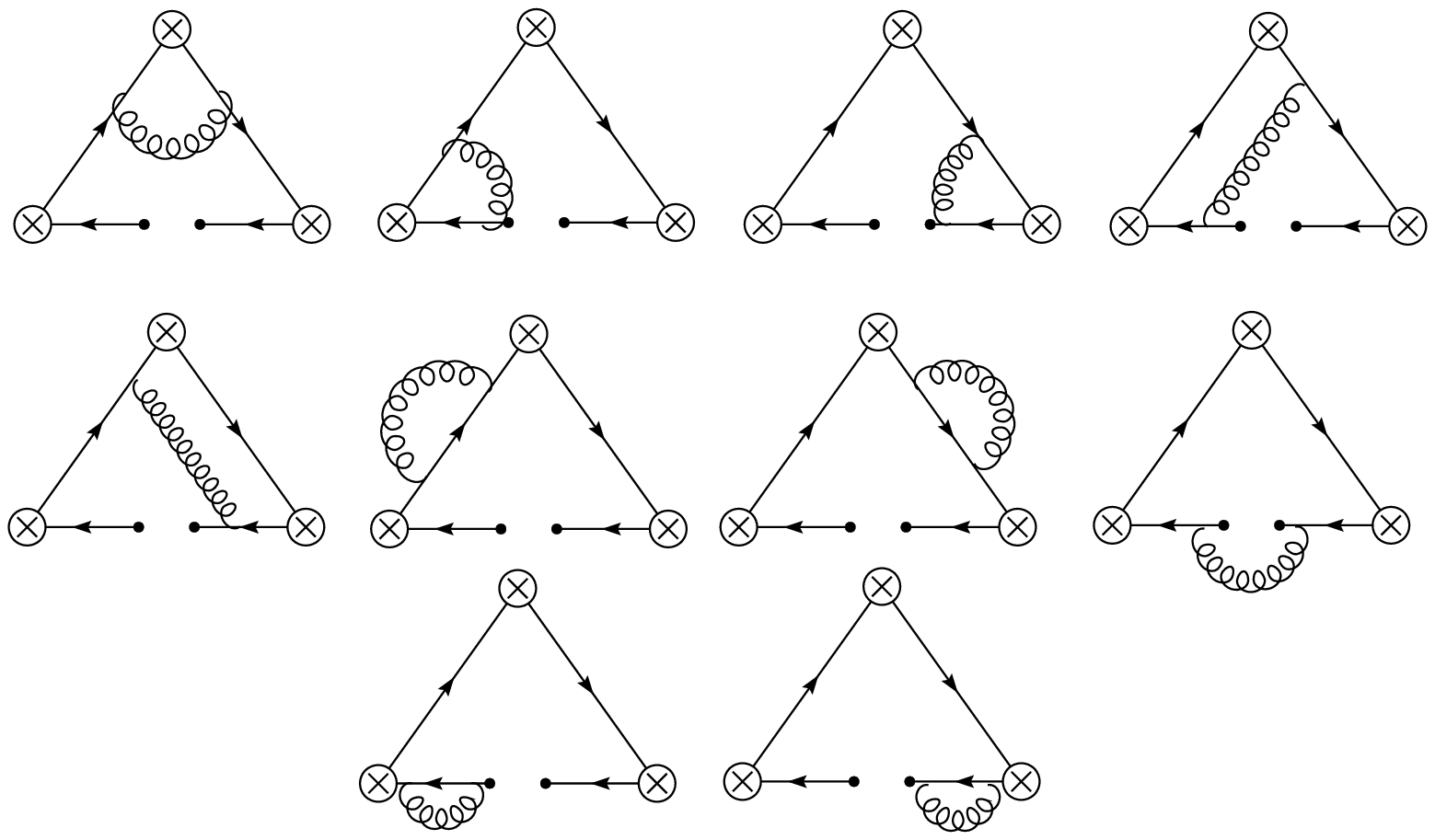}\end{center}\vspace*{-0.9cm}
  \caption{Gluonic corrections to Fig.~\ref{cap:3-point-tree} \label{cap:3-point-gluonic}\vspace*{-0.5cm}}
\end{figure}

\noindent The case of three point functions is of great interest because of
several reasons. First, unlike the two point case, there is a quite
big amount of them that are order parameters of the chiral symmetry breaking. Second, in the framework
of R$\chi$T it involves vertices among resonances 
and so it is useful to learn how they interact. Third, there are a lot of $\chi$PT LEC's
that can be determined with these Green Functions. And fourth, by
means of the LSZ reduction formula we can relate the Green Functions
with form factors entering the calculation of many interesting hadronic observables. Some phenomenological applications of these functions can be found in Ref.~\cite{Narison}. Fig.~\ref{cap:3-point-tree} accounts
for the $\mathcal{O}(\alpha_{s}^{0})$ 
contribution
and Fig.~\ref{cap:3-point-gluonic} represents their first
order gluonic corrections.\vspace*{-0.4cm}

\vspace*{-0.4cm}

Due to the length of the results of this section we will not include explicit expressions, which can be found in Ref.~\cite{mateu-jamin-pich}. These three point functions can be divided into two subsets: odd and even intrinsic parity sector.

The first set contains three Green Functions, namely $\left\langle VVP \right\rangle$,  $\left\langle AAP \right\rangle$ and $\left\langle VAS \right\rangle$. Due to Ward identities and time reversal, the three of them can be written as:
\begin{eqnarray}
\Pi_{\mu\nu}^{abc}(p,q)=\varepsilon_{\mu\nu\alpha\beta}\,p^{\alpha}q^{\beta}g^{abc}\,\Pi\,(p^2,q^2,r^2)\,,
\end{eqnarray} 
where $g^{abc}$  can be either $d^{abc}$ or $f^{abc}$ depending on the time reversal properties of the Green Function. All these functions are finite and do not need renormalization.

In the even intrinsic parity sector we have two groups. $\left\langle SSS \right\rangle$ and $\left\langle SPP \right\rangle$ need to be renormalized; $\left\langle VVS \right\rangle$,  $\left\langle AAS \right\rangle$ and $\left\langle VAP \right\rangle$ are 
finite. Each one of the latter have only two independent Lorentz structures, what defines two form factors.

\section{Renormalization group equations and anomalous dimensions}\noindent
The definitions of the anomalous dimensions of the QCD parameters,
currents and operators are the following:
\begin{eqnarray}
\mu\frac{\mathrm{d}J(\mu)}{\mathrm{d}\mu} & = & -\gamma_{J}J(\mu)\,,\\
\frac{1}{m(\mu)}\mu\frac{\mathrm{d}m(\mu)}{\mathrm{d}\mu} & = & -\gamma_{m}=-\frac{3}{2}C_{F}\frac{\alpha_{s}}{\pi}\,,\\
\frac{1}{\left\langle \bar{q}q\right\rangle (\mu)}\mu\frac{\mathrm{d}\left\langle \bar{q}q\right\rangle (\mu)}{\mathrm{d}\mu} & = & -\gamma_{\left\langle \bar{q}q\right\rangle }\,.\end{eqnarray}
Since $m(\mu)S(\mu)$, $m(\mu)P(\mu)$ and $m(\mu)\left\langle \bar{q}q\right\rangle (\mu)$
are renormalization group invariant objects, then $\gamma_{S}=\gamma_{\left\langle \bar{q}q\right\rangle }=-\gamma_{m}$,
and since $V^{\mu}$ and $A^{\mu}$ currents are conserved $\gamma_{V}=\gamma_{A}=0$.
On the other hand:
\begin{equation}
\mu\frac{\mathrm{d}\left\langle \Pi_{i=1}^nJ_{i}  \right\rangle (\mu)}{\mathrm{d}\mu}=-\left(  \sum_{i=1}^n\gamma_{i} \right)\left\langle \Pi_{i=1}^nJ_{i}\right\rangle (\mu)\label{weinbergformula}\,,\end{equation}
and under the conditions of order parameters in the chiral limit 
eq.~(\ref{weinbergformula}) leads to:
\begin{equation}
\left(\sum_{i=1}^n\gamma_{i}+\gamma_{m}+\mu\frac{\partial}{\partial\mu}\right)\left\langle \Pi_{i=1}^nJ_{i}\right\rangle (\mu)  =  0\,,\label{rge} \end{equation}
which tell us the scale dependence of the Green Functions. The mass anomalous dimension appears in (\ref{rge}) as a consequence of the running of the $\left\langle \bar{q}q\right\rangle $ operator, the only one playing a role (the $\beta$ function is already $\mathcal{O}(\alpha_s^2)$). In particular,
if $\sum_{i=1}^n\gamma_{i}+\gamma_{m}=0$ 
there is
no explicit $\mu$-dependence. All our results follow the renormalization group equations derived presented in this section, which is a good consistency check \cite{mateu-jamin-pich}.

\section{Summary}
\noindent
We have shown that after a correct understanding of the infrared divergences present in 
the order parameter Green Functions of the chiral symmetry breaking those cancel out
. Moreover they are responsible for the running of $\left\langle \bar{q}q \right\rangle $.

We have written renormalization group equations for all our Green Functions 
and our results follow the scale dependence derived from them. 

These results might be used to compute corrections to 
some spectral sum rules analyis so far done at leading order. But its main utility is to understand how to go beyond the minimal hadronic ansatz and how to include $\alpha_s$ corrections and scale dependence to the $\chi$PT LEC's determination by imposing the short distance constraints (now with a logarithmic dependence). Now we understand the problems arising when matching SSS and SPP results of OPE and R$\chi$T \cite{matching}, but still we do not know how to solve them \cite{mateu-jamin-pich}.
\vspace*{0.3cm}

\begin{large}\textbf{Acknowledgements}\end{large}\vspace*{0.2cm}

\noindent This work has been supported in part by
EU FLAVIAnet network (MRTN-CT-2006-035482)
, the Spanish Ministry of Education and Science
(grant FPA2004-00996), Generalitat Valenciana
(GRUPOS03/013 and GV05-164) and by ERDF funds from the EU Commission.

\end{document}